\documentclass[twocolumn,superscriptaddress,floatfix,preprintnumbers,prd ,nofootinbib,hyperref]{revtex4-2}
\usepackage[colorlinks=true,breaklinks=true]{hyperref}
\hypersetup{allcolors=[rgb]{0.0 0.0 0.6},linkcolor=[rgb]{0.75 0.05 0.05}}
\usepackage{mathtools}
\usepackage{bm}
\usepackage{cancel}
\usepackage{physics}
\usepackage[dvipsnames]{xcolor}

\LetLtxMacro{\oldcite}{\cite}
\renewcommand{\cite}[1]{\mbox{\oldcite{#1}}}

\usepackage{setspace}

\usepackage{orcidlink}
\usepackage{slashed}

\long\def\exclude#1{}

\newcommand{\br}{{\bf r}}
\newcommand{\bp}{{\bf p}}
\newcommand{\bq}{{\bf q}}

\newcommand{\sss}{{\hbox{\boldmath$\sigma$}}}

\bibpunct{[}{]}{,}{n}{}{,}

\begin{document}

\title{Collective neutrino-antineutrino oscillations in dense neutrino environments?}

\author{Damiano F.\ G.\ Fiorillo \orcidlink{0000-0003-4927-9850}}
\affiliation{Niels Bohr International Academy, Niels Bohr Institute,
University of Copenhagen, 2100 Copenhagen, Denmark}

\author{Georg G.\ Raffelt
\orcidlink{0000-0002-0199-9560}}
\affiliation{Max-Planck-Institut f\"ur Physik (Werner-Heisenberg-Institut), Boltzmannstr.~8, 85748 Garching, Germany}

\author{G\"unter Sigl
\orcidlink{0000-0002-4396-645X}}
\affiliation{Universit\"at Hamburg, II.~Institut f\"ur Theoretische Physik, 22761 Hamburg, Germany}

\date{4 January 2024}

\begin{abstract}
The paradigm-changing possibility of collective neutrino-antineutrino oscillations was recently advanced in analogy to collective flavor oscillations. However, the amplitude for the backward scattering process $\nu_{\bp_1}\overline\nu_{\bp_2}\to\nu_{\bp_2}\overline\nu_{\bp_1}$ is helicity-suppressed and vanishes for massless neutrinos, implying that there is no off-diagonal refractive index between $\nu$ and $\overline\nu$ of a single flavor of massless neutrinos. For a nonvanishing mass, collective helicity oscillations are possible, representing de-facto $\nu$--$\overline\nu$ oscillations in the Majorana case. However, such phenomena are suppressed by the smallness of neutrino masses as discussed in the previous literature
\end{abstract}

\maketitle

\section{Introduction}

In a series of papers \cite{Sawyer:2023dov, Sawyer:2022ugt}, Sawyer has advanced the potentially paradigm-changing idea that dense neutrino environments could spawn collective neutrino-antineutrino oscillations. Ever since Pantaleone's seminal paper \cite{Pantaleone:1992eq} it has been understood that neutrinos of different flavor can evolve collectively, where the effect originates from a flavor off-diagonal refractive effect that exists in a gas consisting of neutrinos or antineutrinos that are coherent superpositions of different flavors. If we substitute the attribute {\em flavor} with the attribute {\em lepton number}, one may think that a similar effect could obtain in a single-flavor gas of $\nu$ and $\overline\nu$.

In the limit of massless neutrinos and concomitant absence of flavor mixing, the phenomenon of fast flavor conversion still opens the possibility of large flavor coherence building up~\cite{Sawyer:2015dsa, Chakraborty:2016lct, Tamborra:2020cul}. In particular, fast pair conversion of the type $\nu_e\overline{\nu}_e\to \nu_x\overline{\nu}_x$ can obtain without flavor-lepton number violation. The corresponding classical instability (meaning an instability of the mean field) depends on a nontrivial angle distribution, which is why it had been overlooked for a long time until it was recognized by Sawyer~\cite{Sawyer:2015dsa}. The simplest toy model is a homogeneous system with a minimum of three beams~\cite{Padilla-Gay:2021haz, Fiorillo:2023mze, Fiorillo:2023hlk} (see also Ref.~\cite{Dasgupta:2017oko} for a four-beam model), which is easily understood. Two-flavor dynamics corresponds to flavor isospin dynamics and flavor conservation to angular momentum conservation. With only two beams, there are only two flavor spins, total angular momentum is conserved, thus leaving only one dynamical variable and only a simple precession around the conserved angular momentum. With three beams (now each thought of as spins), one retains the possibility of pair-wise flips of the spins without changing total angular momentum.\footnote{Sawyer finds an instability even for two beams, but this is caused by an algebraic error. In Eq.~(4) of Ref.~\cite{Sawyer:2023dov}, the right-hand side of the 3rd and 4th lines should carry a minus sign. For any occupation of his beams, the eigenvalues are real and do not show exponential growth.}

Many years ago, two of us have derived a kinetic equation for flavor-mixed neutrinos \cite{Sigl:1993ctk}. (See Refs.~\cite{Dolgov:1980cq, Rudsky} for other early derivations.) These often-used equations allow for fast flavor conversion, even if the fast flavor instability had not yet been discovered. On the other hand, terms corresponding to the possibility of $\nu$--$\overline\nu$ coherence were left out from the start. Our argument was that such correlations would violate lepton number and thus would not build up if there was no lepton-number violation in the system and the initial state had no such correlations. With hindsight, this argument was incomplete because by the same token one might have excluded flavor conversion without neutrino masses and mixing. We ignored the possibility of a classical instability that might build up such correlations even from a minimal seed. 

In this sense, a similar fast conversion effect could arise in the single-flavor $\nu$--$\overline\nu$ system without violating global lepton-number conservation. If there was a classical instability, correlations could build up, and neutrinos and antineutrinos could be shuffled between different energies and directions with potentially important effects in neutrino-dense environments~\cite{Sawyer:2023dov, Sawyer:2022ugt}. Such effects would require off-diagonal refraction, i.e., a suitably prepared background medium needs to mix $\nu$ with $\overline\nu$. However, the left-handedness of weak interactions implies that in the massless limit, helicity plays the role of lepton number, which is why it is so hard to distinguish Dirac from Majorana neutrinos---observable differences disappear with vanishing neutrino mass. By the same token, a $\nu$--$\overline\nu$ transition requires a helicity flip. It is helicity playing the role of lepton number that nixes the analogy to the flavor case as we will see.

Some of these questions were already addressed by several groups a few years ago, framed as the question of helicity oscillations, always finding that these effects are suppressed by the smallness of neutrino mass \cite{Vlasenko:2013fja, Cirigliano:2014aoa, Vlasenko:2014bva, Dobrynina:2016rwy, Kartavtsev:2015eva, Volpe:2013uxl, Serreau:2014cfa, Chatelain:2016xva}. For the Majorana case, these studies correspond to the question of $\nu$--$\overline\nu$ transitions in a polarized neutrino background. As we will see, the same conclusion pertains to the Dirac case, in the sense that, in the massless limit, no $\nu$--$\overline\nu$ transition can be prompted by a polarized neutrino background.

\section{Off-Diagonal Refraction}

\begin{figure}[t]
    \centering
    \includegraphics[width=0.65\columnwidth]{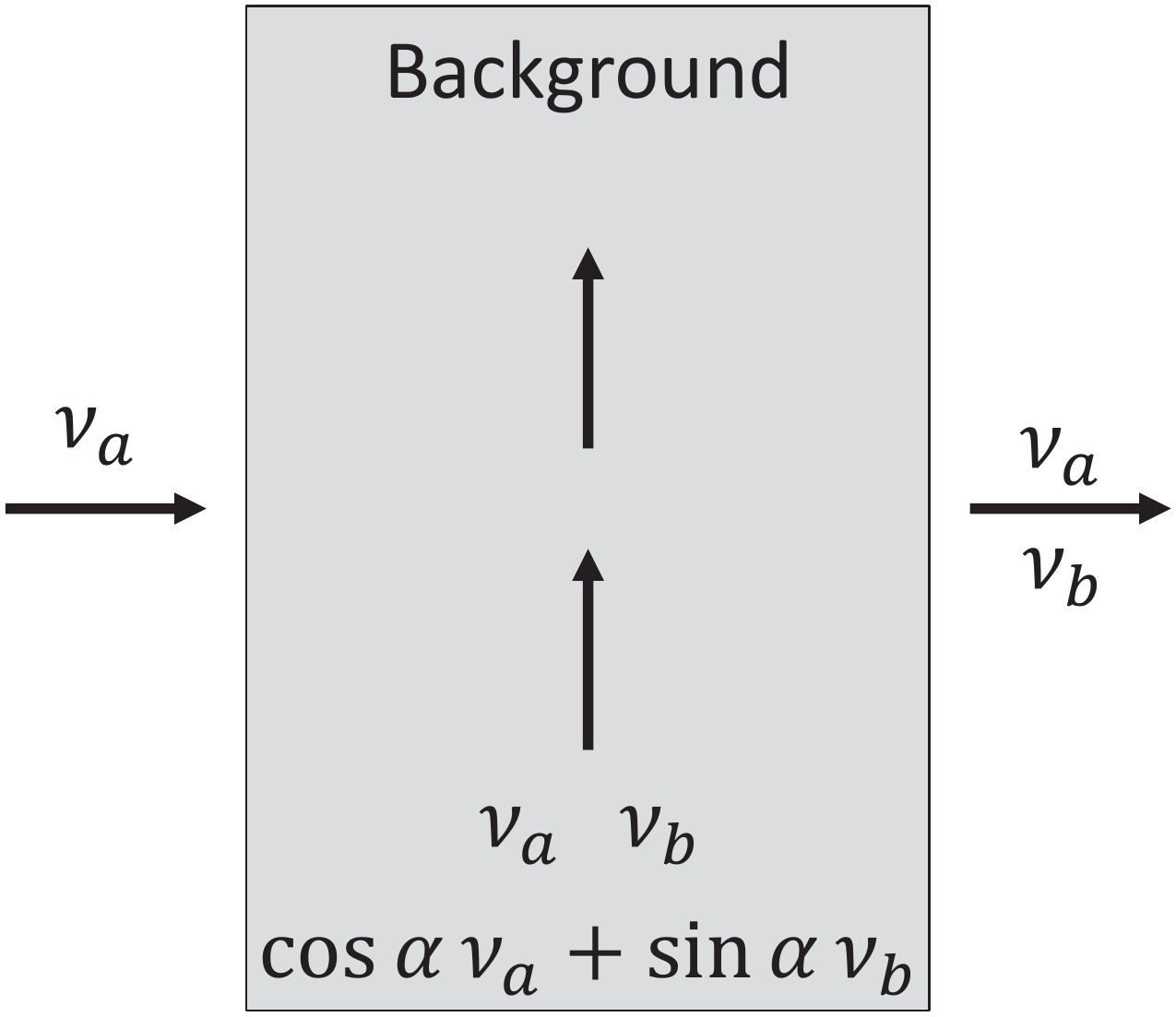}
    \vskip6pt
    \caption{Test neutrinos with attribute $a$ crossing a background of neutrinos with a statistical mixture of attributes $a$ and $b$, or a coherent superposition. (Setup and figure inspired by Friedland and Lunardini \cite{Friedland:2003dv}.)
    }
    \label{fig:beams}
    \vskip6pt
\end{figure}

To show this suppression in the most direct way, we are inspired by Friedland and Lunardini \cite{Friedland:2003dv} to frame the question as illustrated in Fig.~\ref{fig:beams}. We consider a test neutrino with flavor $a$ crossing a background beam and ask under which circumstances it emerges in a different flavor state $b$? If the background consists of a statistical mixture of neutrinos with flavors $a$ or $b$, it is possible that the test neutrino collides with a $\nu_b$ and they can exchange momentum. This would happen with a rate proportional to $G_{\rm F}^2$ times the density of background neutrinos, which we call a ``hard collision.'' For a coherent effect on the refractive level, proportional to $G_{\rm F}$, many such elementary events need to add coherently as explained by Friedland and Lunardini. \textit{However, no refractive effect obtains if there is no elementary hard scattering process that could produce the same result of a $\nu_b$ emerging with the same momentum as the original $\nu_a$.} In addition, the background neutrinos need to be in a coherent superposition $\nu_x=\cos\alpha\,\nu_a+\sin\alpha\,\nu_b$, producing a flavor off-diagonal refractive effect. The test $\nu_a$ slowly develops a $\nu_b$ component and emerges as a coherent $\nu_a$--$\nu_b$ superposition. Whereas in the hard-collision case, the modified flavor comes at the expense of one specific background neutrino, it here derives from the entire medium that changes its own coherent flavor state by a small amount. The entire medium ``recoils'' in flavor space, not a single neutrino of the medium.

To see this effect more explicitly we observe that the refractive potential of a neutrino in a background of its own flavor is twice that in a background of a different flavor, owing to the final-state exchange amplitude that arises for identical flavors. If the background is prepared in the coherent superposition $\nu_x=\cos\alpha\,\nu_a+\sin\alpha\, \nu_b$ we may expand the test neutrino $\nu_a$ in terms of $\nu_x$ and the orthogonal flavor mixture $\nu_y$, with $\nu_x$ feeling twice the refractive effect of $\nu_y$, so these two propagation eigenstates develop a phase difference, leading to the appearance of a $\nu_b$ component of the original $\nu_a$. 

If the background consists of antineutrinos and the test particle is still a neutrino, the outcome is analogous if the background medium is in a coherent superposition $\overline\nu_x=\cos\alpha\,\overline\nu_a+\sin\alpha\,\overline\nu_b$. Once more we can expand the test $\nu_a$ in its $\nu_x$ and $\nu_y$ components. The forward-scattering amplitude of a $\nu$ in a $\overline\nu$ bath of its own flavor is twice that of a different flavor, although the factor of~2 here arises not from an exchange amplitude, but from the option of $\nu\overline\nu\to\nu\overline\nu$ proceeding through $Z^0$ exchange (\hbox{$t$-channel}) or through the $s$-channel $\nu\overline\nu\to Z^0\to\nu\overline\nu$. The elementary process that causes the refractive effect is the $s$-channel $\nu_a\overline\nu_a\to Z^0\to\nu_b\overline\nu_b$ that allows a $\nu_b$ to emerge with the momentum of the original $\nu_a$. The conclusion is the same as for a background of neutrinos, due to the crossing symmetry which connects the $s$-channel process with the $u$-channel $\nu\nu\to\nu\nu$ process. For a coherent refractive effect to appear, once more it is not enough for the background medium to contain $\overline\nu_a$, but we need a coherent superposition of both flavors.

If the background consists of neutrinos, the largest hard-collision effect obtains if the background is purely $\nu_b$, whereas if it consists of antineutrinos, the incoherent effect is largest for a pure $\overline\nu_a$ background. In both cases, the largest refractive effect obtains if the background is an equal superposition of both flavors. All of this follows from the usual equations or from the detailed microscopic discussion of Friedland and Lunardini \cite{Friedland:2003dv}.

The situation is different if we consider active-sterile conversion, where $a$ stands for an active flavor and $b$ for a hypothetical sterile one. If the background is once more a statistical mixture, $\nu_a$ cannot emerge as $\nu_b$ because the scattering cross section with a sterile neutrino vanishes. Even if the background has been prepared in a coherent superposition, there is no off-diagonal refractive effect, the test neutrino cannot emerge in the sterile flavor by coherent or incoherent interactions. Of course it could oscillate driven by active-sterile vacuum mixing and a mass difference, but this is a different effect.

\section{Neutrino-Antineutrino Backward Scattering}

We now turn to our real question, what happens if the attributes $a$ and $b$ stand for $\nu$ and $\overline\nu$ of one flavor. Let us again consider a statistical mixture of $\nu$ and $\overline\nu$ in the background. Can $\nu$ emerge as $\overline\nu$ after a hard collision? Certainly $\nu$ and $\overline\nu$ have a nonvanishing scattering cross section. However, our question is narrower because we want the participants to exchange momenta (effectively backward scattering), so that $\overline\nu$ emerges with the momentum of the original $\nu$. However, the cross section for such a collision vanishes for massless neutrinos.

\begin{figure}[t]
    \centering
    \includegraphics[width=1.0\columnwidth]{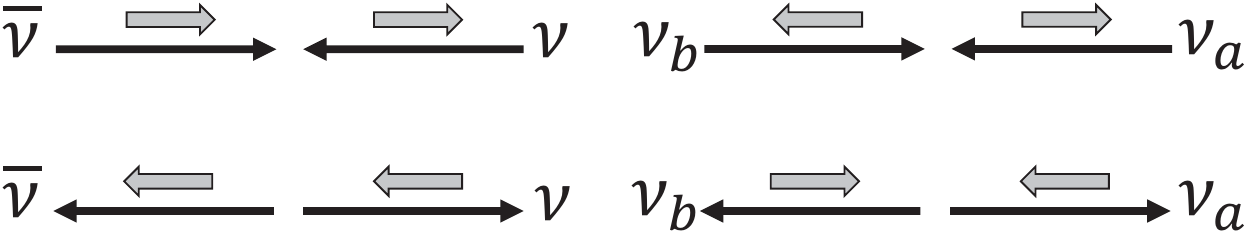}
    \caption{CM-frame neutrino scattering with momentum exchange (backward scattering), with grey arrows indicating their spin, initial state top, final state bottom. {\em Left:\/} $\nu$--$\overline\nu$ scattering, which has vanishing amplitude because the total angular momentum would have to be flipped even though it is conserved. {\em Right:\/} Left-handed neutrinos of different flavor, which has non-zero amplitude. The total angular momentum vanishes before and after the collision.
    }
    \label{fig:uchannel}
\end{figure}

This is seen most easily if we observe that the interaction $\nu(\bp_1)+\overline\nu(\bp_2)\to\nu(\bp_2)+\overline\nu(\bp_1)$ can be viewed in the CM frame instead of the lab frame, so we may assume $\bp_1=-\bp_2$ and the two particles collide head on. For vanishing mass, helicity is invariant under Lorentz transformations and so they must have opposite helicity, meaning equal spin along the beam direction (Fig.~\ref{fig:uchannel}). After an exchange of momenta, their spins must be flipped to recover left-chiral states. However, angular momentum along the beam direction is conserved, so the amplitude for this process must vanish.

The vanishing of this amplitude  implies that there is no elementary process to produce a $\overline\nu$ moving in the same direction as the original $\nu$ and consequently, there is also no refractive effect of this type. While our simple helicity argument explains directly why this amplitude must vanish, the same can be shown more formally in the lab frame (see Appendix~\ref{sec:uchannel}).

\section{Neutrino-Antineutrino Refraction}

A test $\nu$ in a bath of same-flavor $\nu$ feels the usual weak potential, whereas in a bath of $\overline\nu$ it feels the same potential with opposite sign, all of which is the usual neutrino-neutrino refractive effect. Therefore, if we prepare a test beam in an equal coherent superposition of $\nu$ and $\overline\nu$, this amounts to linear polarization because $\nu$ and $\overline{\nu}$ correspond to circular polarization, the usual helicity states. If this linearly polarized test beam propagates in a medium of pure $\nu$, the two circular polarization components develop different phases and hence the linear polarization turns along the test beam in analogy to a linearly polarized photon beam in a birefringent medium. This could be a sugar solution which has handedness built into the molecules, allowing the isotropic medium to distinguish between the two photon helicities. The same happens here for neutrinos. Therefore, the chirality of a neutrino medium produces an analogous birefringence effect.

On the other hand, in the previous section we have concluded that in the opposite situation of a $\nu$ propagating in a background beam of, say, linearly polarized neutrinos (an equal coherent superposition of $\nu$ and $\overline\nu$) will propagate as in vacuum. It is blind to the coherence of the linearly polarized background and can only see the sum of the potentials produced by the populations of background $\nu$ and $\overline\nu$, irrespective of this being a statistical mixture or a coherent superposition.

There is no contradiction in this seemingly asymmetric behavior. Both neutrinos and antineutrinos have a nonvanishing forward-scattering amplitude on the background states, the amplitudes are different, and so the two components develop a phase difference. In the opposite case of a $\nu$ propagating in a nontrivially polarized background, there is no elementary amplitude that would allow it to flip its helicity. We make the absence of such an off-diagonal refractive component more formally explicit in Appendix~\ref{sec:refraction}.

\section{Conclusions}

We have presented simple arguments to explain that $\nu$--$\overline\nu$ coherence cannot build up, in contrast to fast flavor oscillations, not because of lepton-number conservation, but because of the helicity structure of the $u$-channel amplitude of $\nu$--$\overline\nu$ forward scattering. This conclusion is opposite to the one reached by Sawyer \cite{Sawyer:2022ugt, Sawyer:2023dov}, the difference arising from an incorrect $u$-spinor in his Dirac algebra (see Appendix~\ref{sec:uchannel}). 

As a more general comment, a neutrino gas initially consisting of an equal mixture of $\nu_e$ and $\overline\nu_e$ will chemically equilibrate by normal collisions (at the order $G_{\rm F}^2$) to a three-flavor mixture. For appropriate angular distributions that engender a fast flavor instability, this process will be accelerated on the refractive level (order $G_{\rm F}$), but the phase-space restrictions of forward scattering will not easily allow for true equilibration, although some form of flavor equipartition may obtain, see, e.g., Refs.~\cite{Zaizen:2022cik, Nagakura:2022kic, Padilla-Gay:2022wck, Zaizen:2023ihz, Johns:2023jjt, Cornelius:2023eop, Fiorillo:2023ajs,Johns:2023xae}. If the initial mixture consists of $\nu_e$ and $\nu_x$ with distinct energy distributions and a suitable angle distribution, once more fast flavor instabilities can accelerate the equipartition among energies that eventually arises anyway on the collisional level, i.e., fast flavor instabilities are not limited to the simplest example of pair conversion.

Likewise, any energy and angle distribution of massless $\nu$ and $\bar\nu$ will preserve lepton number due to the left-handedness of weak interactions where helicity plays the role of lepton number. Collisions will equilibrate this distribution among energies. The main difference to the fast flavor case is that this process cannot be accelerated by  refractive effects.

For neutrinos with nonzero mass, collisions populate the sterile neutrino helicities in the Dirac case, and equilibrate $\nu$ with $\bar\nu$ in the Majorana case. Both effects can also arise on the refractive level, which in the Majorana case amounts to $\nu$--$\overline\nu$ oscillations. These questions were investigated a few years ago by several groups \cite{Vlasenko:2013fja, Cirigliano:2014aoa, Vlasenko:2014bva, Dobrynina:2016rwy, Kartavtsev:2015eva, Volpe:2013uxl, Serreau:2014cfa, Chatelain:2016xva}, and it was found that they involve a neutrino mass factor, in concordance with the pedestrian arguments presented here. Therefore, a ``fast flavor'' acceleration is not possible for either the Dirac or Majorana case. The rate of collective neutrino-antineutrino oscillations must always vanish with vanishing neutrino mass.

\section*{Acknowledgments}

We acknowledge a correspondence with Ray Sawyer, which however did not lead to convergent views. We thank Alexander Kartavtsev for discussions, and Basu Dasgupta, Irene Tamborra, and Shashank Shalgar for insightful comments and suggestions on the manuscript. DFGF is supported by the Villum Fonden under Project No.\ 29388 and the European Union's Horizon 2020 Research and Innovation Program under the Marie Sk{\l}odowska-Curie Grant Agreement No.\ 847523 ``INTERACTIONS.'' GGR acknowledges partial support by the German Research Foundation (DFG) through the Collaborative Research Centre ``Neutrinos and Dark Matter in Astro- and Particle Physics (NDM),'' Grant SFB-1258-283604770, and under Germany’s Excellence Strategy through the Cluster of Excellence ORIGINS EXC-2094-390783311. GS acknowledges support by the Deutsche Forschungsgemeinschaft (DFG, German Research Foundation) under Germany’s Excellence Strategy -- EXC 2121 ``Quantum Universe'' -- 390833306. 

\appendix

\section{Amplitude for \texorpdfstring{\boldmath{$\nu(\bp_1)+\overline\nu(\bp_2)\to\nu(\bp_2)+\overline\nu(\bp_1)$}}{}
}
\label{sec:uchannel}

\begin{figure}[t]
    \centering
    \includegraphics[width=1.0\columnwidth]{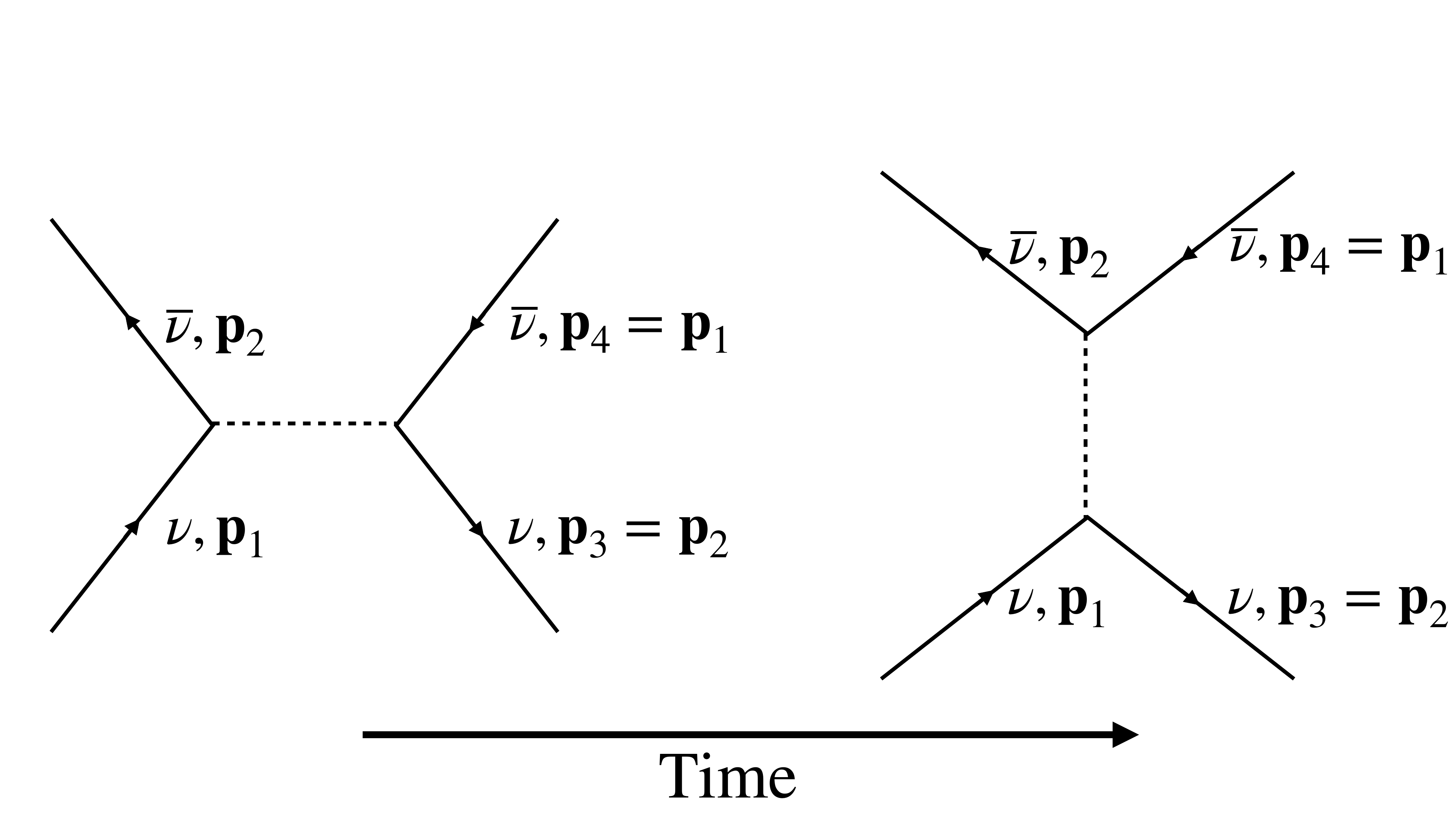}
    \caption{$s$-channel (left) and $t$-channel (right) process for neutrino-antineutrino scattering with momentum exchange. The amplitude for both diagrams vanishes identically for any values of $\bp_1$ and $\bp_2$ (see text).}
    \label{fig:diagrams_exchange}
\end{figure}

In the main text we have argued on elementary grounds that the amplitude for the $u$-channel forward scattering process (exact backward scattering) $\nu(\bp_1)+\overline\nu(\bp_2)\to\nu(\bp_2)+\overline\nu(\bp_1)$ must vanish. The same result can be found more explicitly in the lab frame. The $t$-channel forward-scattering amplitude (Fig.~\ref{fig:diagrams_exchange}) has the structure
\begin{equation}\label{eq:matrixelement}
    \bar u_{\bp_2} \gamma^\mu u_{\bp_1} ~\bar v_{\bp_2} \gamma_\mu v_{\bp_1},
\end{equation}
where $u_{\bp}$ is the spinor for a neutrino with momentum $\bp$ and negative helicity, $v_{\bp}$ the one for an antineutrino with positive helicity. To simplify notation, we do not explicitly show helicities because for our massless limit, it is consistent to assume a fixed chirality for all participating states. As the spinors are already taken for chiral neutrinos, the left-handed projection is not needed in the matrix element. In a field operator they would appear in the combination 
\begin{equation}
    a_{{\bf p}}u_{{\bf p}}
  e^{-iE_{\bf p}t+i{\bf p}\cdot{\bf r}}+b^\dagger_{{\bf p}}v_{{\bf p}}
  e^{iE_{\bf p}t-i{\bf p}\cdot{\bf r}}
\end{equation}
with $a_{{\bf p}}$ the annihilator of a helicity-minus neutrino of physical momentum ${\bf p}$ and \smash{$b^\dagger_{{\bf p}}$} the creator of a helicity-plus anti-neutrino of physical momentum $-{\bf p}$. Once more, we do not explicitly show the helicities.

For neutrino-antineutrino scattering, in addition there is an $s$-channel amplitude with the structure
\begin{equation}\label{eq:matrixelements}
    \bar u_{\bp_2} \gamma^\mu v_{\bp_1} ~\bar v_{\bp_2} \gamma_\mu u_{\bp_1}.
\end{equation}
For comparison, in the case of neutrino-neutrino interaction, leading to the fast flavor instability, the forward scattering amplitude corresponding to the $t$-channel is
\begin{equation}\label{eq:matrixelementt}
    \bar u_{\bp_1} \gamma^\mu u_{\bp_2} ~\bar u_{\bp_2} \gamma_\mu u_{\bp_1}.
\end{equation}
An amplitude with the same structure also describes $\overline\nu$--$\overline\nu$ scattering. This amplitude does not vanish, corresponding to the right panel in Fig.~\ref{fig:uchannel}.

A key observation, perhaps at first surprising, is that the $u$-spinor for a helicity-minus neutrino and the $v$-spinor for a helicity-plus antineutrino, both moving in the same direction, are the same, up to an arbitrary phase. It is less surprising once we realize that, for zero mass, both spinors obey the same equation $\slashed{p}u_{\bp,-}=\slashed{p}v_{\bp,+}=0$ augmented by the helicity condition $(1+\gamma_5)u_{\bp,-}=(1+\gamma_5)v_{\bp,+}=0$, and therefore are identical. (We here indicate explicitly the helicity by the indices $\pm$.) Notice that this equality directly implies that the $t$-channel and the $s$-channel spinor amplitudes are identical; therefore, since the total amplitude is the difference between them, where the minus sign arises from the exchange of two fermions, one can already conclude that the total amplitude vanishes. However, one can prove the stronger result that both amplitudes vanish separately.

A direct way is to use the Fierz identities, by which one directly obtains
\begin{equation}
 \bar u_{\bp_2} \gamma^\mu u_{\bp_1} ~\bar v_{\bp_2} \gamma_\mu v_{\bp_1}
    =-\bar u_{\bp_2} \gamma^\mu v_{\bp_1} ~\bar v_{\bp_2} \gamma_\mu u_{\bp_1}.   
\end{equation}
Since the two amplitudes must also be equal to each other, they must both vanish. The same argument applied to the $t$-channel exchange for two neutrinos does not lead to the same conclusion, since the exchange of the two spinors actually leads to the nontrivial result
\begin{equation}
\bar u_{\bp_1} \gamma^\mu u_{\bp_2} ~\bar u_{\bp_2} \gamma_\mu u_{\bp_1}
=-\bar u_{\bp_1} \gamma^\mu u_{\bp_1} ~\bar u_{\bp_2} \gamma_\mu u_{\bp_2}=-4p_1\cdot p_2,
\end{equation}
which can be directly confirmed by the usual trace rules.

One can obtain the same result in a pedestrian way by writing it out explicitly. The massless spinors in the Dirac representation are found, for example, in Eq.~(4.67) of the textbook~\cite{Thomson:2013zua}, where we include a factor $1/\sqrt2$
\begin{equation}
    u_{\bp,-}=v_{\bp,+}=\sqrt{\frac{|\bp|}{2}}
    \begin{pmatrix}
        -\sin\frac{\theta}{2}\\[1.5ex]
        e^{i\phi}\cos\frac{\theta}{2}\\[1.5ex]
        \sin\frac{\theta}{2}\\[1.5ex]
        -e^{i\phi}\cos\frac{\theta}{2}
    \end{pmatrix},
\end{equation}
showing the helicity indices explicitly. Here $\theta$ and $\phi$ are the polar coordinates of $\bp$ in some chosen frame of reference. One finds explicitly, for example, $\bar u_{\bp} \gamma^\mu u_{\bp}=(|\bp|,\bp)$ or that Eq.~\eqref{eq:matrixelement} vanishes. If the particles move along the positive $z$-direction, these spinors are
\begin{equation}
    u_{\bp,-}=v_{\bp,+}=\sqrt{\frac{|\bp|}{2}}
    \begin{pmatrix}
        0\\
        1\\
        0\\
        -1
    \end{pmatrix}.
\end{equation}
Notice that this is consistent with the fact that the helicity operator for particles is diag$(\sss,\sss)\cdot\bp$ whereas for antiparticles it is $-$diag$(\sss,\sss)\cdot\bp$, where $\sss$ are the Pauli matrices.
While for massless neutrinos it would make more sense to use the chiral representation, we used the Dirac representation to  compare directly with the unnumbered equation after Eq.~(17) in Ref.~\cite{Sawyer:2022ugt} and see that the expression for their $u$-spinor is not correct.

That Sawyer's expression for the $u$-spinor must be wrong is seen from his expression for the current, consisting of $\bar u\gamma^0 u=+1$ and $\bar u\gamma^3 u=-1$, but the current for a particle moving in the positive $z$-direction must have a positive 3-component. His $u$-spinor is for a particle moving in the negative $z$-direction.

\section{Refractive index from neutrino-antineutrino coherence}
\label{sec:refraction}

In the main text we have argued on elementary grounds that a $\nu$ propagating in a background beam which is nontrivially polarized remains blind to this coherence and does not see an off-diagonal refractive index, i.e., such a medium never mixes $\nu$ with $\overline\nu$. To prove more formally that there is actually no such off-diagonal refractive effect it is instructive to explicitly compute the refractive energy shift felt by a neutrino in the background with a coherent superposition of neutrinos and antineutrinos. To fully characterize the background, we introduce an extended density operator
\begin{equation}
    \mathcal{D}^{\alpha\beta}_\bp=\begin{pmatrix}
         a^\dagger_{\beta\bp}a_{\alpha\bp}&& a^\dagger_{\beta\bp}b_{\alpha\bp}\\  b^\dagger_{\alpha\bp}a_{\beta\bp} &&  b^\dagger_{\alpha\bp}b_{\beta\bp}
    \end{pmatrix},
\end{equation}
where $\alpha$ and $\beta$ are flavor indices. The corresponding density matrix is
\begin{equation}
    \rho^{\alpha\beta}_{\bp}=\left\langle\mathcal{D}^{\alpha\beta}_\bp\right\rangle    
    =\begin{pmatrix}
        n^{\alpha\beta}_\bp && s^{\alpha\beta}_\bp \\[1ex]
        s^{\alpha\beta*}_\bp && \overline{n}^{\alpha\beta}_\bp
    \end{pmatrix}.
\end{equation}

To determine the refractive energy shift, we start from the interaction Hamiltonian of neutrinos, which we write
\begin{equation}
    \mathcal{V}=\frac{\sqrt{2}G_{\rm F}}{2}\int d^3 \br \sum_{\alpha\beta}\overline{\nu}_\alpha(\br)\gamma^\mu \nu_\alpha(\br) \overline{\nu}_\beta(\br)\gamma_\mu \nu_\beta(\br).
\end{equation}
Expanding the field operators as
\begin{equation}
    \nu_\alpha(\br)=\int \frac{d^3\bp}{(2\pi)^3}\left[a_{\alpha\bp}u_{\bp}e^{i\bp\cdot\br}+b^\dagger_{\alpha\bp}v_{\bp}e^{-i\bp\cdot \br}\right],
\end{equation}
where the spinors are normalized by $\overline{u}_{\bp}\gamma^0 u_{\bp}=\overline{v}_{\bp}\gamma^0v_{\bp}=1$, we can obtain all the different terms of interaction among $\nu$ and $\overline\nu$. For the purposes of $\nu$--$\overline\nu$ coherence, we are interested in those terms that annihilate and create a particle and an antiparticle. Extracting the relevant terms from the expansion, we obtain
\begin{widetext}
\begin{equation}
    \mathcal{V}^{\overline{\nu}\nu}=\sqrt{2}G_{\rm F}\sum_{\left\{\bp\right\},\alpha,\beta}\left[b_{\alpha\bp_3}a_{\alpha\bp_2}a^\dagger_{\beta\bp_1}b^\dagger_{\beta\bp_4}\overline{v}_{\bp_3}\gamma^\mu u_{\bp_2}\overline{u}_{\bp_1}\gamma_\mu v_{\bp_4}
    +a^\dagger_{\alpha\bp_1}a_{\alpha\bp_2}b_{\beta\bp_3}b^\dagger_{\beta\bp_4}\overline{u}_{\bp_1}\gamma^\mu u_{\bp_2}\overline{v}_{\bp_3}\gamma_\mu v_{\bp_4}\right],
\end{equation}
where $\sum_{\left\{\bp\right\}}$ denotes integration over the phase-space of all four-momenta subject to momentum conservation $\bp_1+\bp_2=\bp_3+\bp_4$. Using the equality of particle-antiparticle spinors, and the Fierz identities, this can be rewritten as
    \begin{equation}
        \mathcal{V}^{\overline{\nu}\nu}=\sqrt{2}G_{\rm F}\sum_{\left\{\bp\right\},\alpha,\beta}\overline{u}_{\bp_1}\gamma^\mu u_{\bp_2}\overline{v}_{\bp_3}\gamma_\mu v_{\bp_4}\left(a^\dagger_{\alpha\bp_1}a_{\alpha\bp_2}b_{\beta\bp_3}b^\dagger_{\beta\bp_4}-b_{\alpha\bp_3}a_{\alpha\bp_2}a^\dagger_{\beta\bp_1}b^\dagger_{\beta\bp_4}\right).
    \end{equation}
Finally, using the anticommutation of different operators, we can bring this term to the normal ordering, where the destruction operators are on the right, obtaining
\begin{equation}
        \mathcal{V}^{\overline{\nu}\nu}=\sqrt{2}G_{\rm F}\sum_{\left\{\bp\right\},\alpha,\beta}\overline{u}_{\bp_1}\gamma^\mu u_{\bp_2}\overline{v}_{\bp_3}\gamma_\mu v_{\bp_4}\left(-a^\dagger_{\alpha\bp_1}a_{\alpha\bp_2}b^\dagger_{\beta\bp_4}b_{\beta\bp_3}-a^\dagger_{\beta\bp_1}a_{\alpha\bp_2}b^\dagger_{\beta\bp_4}b_{\alpha\bp_3}\right).
\end{equation}
\end{widetext}
    
To obtain the refractive energy shift, we can now use the Hartee-Fock procedure and expand $\mathcal{D}_\bp\simeq \rho_\bp+\delta\mathcal{D}_\bp$, where $\delta\mathcal{D}_\bp$ is the operator fluctuation over the mean-field value. The mean field, or Hartree-Fock, approximation corresponds to neglecting the terms quadratic in the fluctuation. Since we are really interested in the off-diagonal refractive index, we can consider only those terms corresponding to the appearance of $s^{\alpha\beta}_\bp$. We can look at only one such term, e.g.\ in the string of operators $a^\dagger_{\beta\bp_1}a_{\alpha\bp_2}b^\dagger_{\beta\bp_4}b_{\alpha\bp_3}$, where the replacement leads to terms of the form $-s^{\beta\alpha*}_\bq a^\dagger_{\beta\bp}b_{\alpha\bp}$, forcing the equality $\bp_1=\bp_3=\bp$, $\bp_2=\bp_4=\bq$; replacing this equality in the spinor matrix element, one recovers the same amplitude which we earlier showed to vanish. This can be seen to happen for all terms, implying that the refractive energy shift induced by a non-vanishing $s^{\alpha\beta}_\bp$ does indeed vanish for massless neutrinos.

Following this strategy also for the other terms, one can directly obtain the effective Hamiltonian felt by a test neutrino which passes through a background medium of other neutrinos. Considering a single flavor, and a neutrino propagating in a medium flowing orthogonally to it (so that the angle between the neutrino direction and the medium velocity is $\cos\theta=0$), and reverting to the bra-ket notation, this effective Hamiltonian is
\begin{equation}
    H=2\sqrt{2}G_F(n_\nu-\overline{n}_\nu)(\ket{\nu}\bra{\nu}-\ket{\overline{\nu}}\bra{\overline{\nu}}).
\end{equation}
As we can see, the term proportional to $s_\nu$ disappears because of the amplitude suppression.

A $\nu$ or a $\overline\nu$ propagates in a background medium consisting of $\nu$ and $\overline\nu$ as if these were a statistical mixture, whether or not the background particles are in a $\nu$--$\overline\nu$ coherent state.

\bibliographystyle{bibi}
\bibliography{References}

\end{document}